# FIRST DETERMINATION OF THE TEMPERATURE OF A LUNAR IMPACT FLASH AND ITS EVOLUTION

**José M. Madiedo[1], José L. Ortiz[2], Nicolás Morales[2]**

[1] Facultad de Ciencias Experimentales, Universidad de Huelva. 21071 Huelva (Spain).
[2] Instituto de Astrofísica de Andalucía, CSIC, Apt. 3004, Camino Bajo de Huetor 50, 18080 Granada, Spain.

**ABSTRACT**

We report the first analysis of a flash produced by the impact of a meteoroid on the lunar surface and recorded both in the near-infrared and in the visible. Despite the fact that similar data have been recently published by other team during the refereeing process of our manuscript (Bonanos et al. 2018), our result still forms the first measurement of the temperature of a telescopic lunar impact flash (Madiedo and Ortiz 2016, 2018). The flash exhibited a peak magnitude of 5.1 ± 0.3 in the near-infrared I band and 7.3 ± 0.2 in the visible, and the total duration of the event in these bands was 0.20 s and 0.18 s, respectively. The origin of the meteoroid was investigated, and we inferred that the most likely scenario is that the impactor that belonged to the sporadic background. The analysis of this event has provided for the first time an estimation of the emission efficiency in the near-infrared $\eta_I$ for sporadic meteoroids impacting the Moon. We have determined that this efficiency is around 56% higher than in the visible band and we have found a maximum impact plume temperature of ~4000 K at the initial phase followed by temperatures of around 3200 K after the peak brightness. The size of the crater produced as a consequence of this impact is also calculated.

**KEYWORDS:** Impact processes, impact flash, Moon, meteoroids, meteors

## 1. INTRODUCTION

Different researchers have studied the impacts of meteoroids on the lunar surface by analyzing the flashes produced during these collisions (Ortiz et al. 1999; Bellot Rubio et al. 2000a,b; Ortiz et al. 2000; Yanagisawa and Kisaichi 2002; Cudnik et al. 2002; Ortiz et al. 2002; Yanagisawa et al. 2006; Cooke et al. 2006; Ortiz et al. 2006; Swift et al. 2011; Madiedo et al. 2014a, Suggs et al. 2014). In this way different parameters can be determined, such as the energy of the impactor, its mass, and the size of the resulting crater. From the frequency of these events, paramount information related to the impact hazard for Earth can be also derived (Ortiz et al. 2006; Madiedo et al. 2014a,b; Suggs et al. 2014). The lunar impact flash monitoring technique has the advantage that a single detector covers a much larger area (typically with an order of magnitude of $10^6$ km$^2$) than that monitored by ground-based systems that analyze meteor and fireball activity in the Earth's atmosphere. However, the results derived from this method depend strongly on the value adopted for the so-called luminous efficiency. This parameter is the fraction of the kinetic energy of the impactor that is emitted as visible light during the impact.

In the last decades, flashes produced by the collision of shower and sporadic meteoroids have been identified in the framework of several monitoring programmes by means of small telescopes and high-sensitivity CCD cameras (Ortiz et al. 2000; Yanagisawa and Kisaichi 2002; Yanagisawa et al. 2006; Cooke et al. 2006; Ortiz et al. 2006; Ortiz et al. 2006; Madiedo et al. 2014a; Suggs et al. 2014; Madiedo et al. 2015a,b). However, the detection of these impact flashes has so far been performed in the visible range. Our team started in 2009 a lunar impact flashes monitoring program named MIDAS (Madiedo et al. 2010; Madiedo et al. 2015a,b), which is the continuation of the lunar impact flashes monitoring project started by the second author in 1999 (Ortiz et al. 1999). Since 2015, in addition to the observations performed in visible band, we are also conducting a monitoring of the night side of the Moon in the near-infrared (NIR) by using a specific NIR filter. In this way, we can analyze the behaviour of these impact flashes in different spectral bands. By following this approach, MIDAS became the first system that can determine the temperature of these impact flashes (Madiedo and Ortiz 2016, 2018). The usefulness of performing observations in the near-infrared was addressed in Cudnik (2010). In this work we focus on a lunar impact flash identified by several of our telescopes on 2015 March 25. It was simultaneously recorded in both visible and NIR bands.

The analysis of this event has provided an estimation of the emission efficiency for flashes produced by sporadic sources in the NIR. Here we use the term emission efficiency in the near Infrared ($\eta_I$) to distinguish it from the luminous efficiency concept ($\eta$), which is applicable to the whole CCD sensitivity wavelength range of 400nm to 900 nm as defined in the initial papers on lunar impact flashes (Ortiz et al. 1999, Bellot et al. 2000a).

## 2. INSTRUMENTATION AND METHODS

The event analyzed here was recorded from our observatory in Sevilla (latitude: 37.34611 ºN, longitude: 5.98055 ºW, height: 23 m above the sea level), during a lunar monitoring campaign conducted on 2015 March 25. Three f/10 Schmidt-Cassegrain telescopes manufactured by Celestron and with diameters of 0.36, 0.28 and 0.24 m were employed. Each telescope used a high-sensitivity CCD video camera (model 902H Ultimate, manufactured by Watec Corporation, Japan). These devices generate an interlaced black and white analogue video signal at a rate of 25 frames per second. This signal is digitized and stored on the hard disk of a PC computer under AVI video files. These video files have a resolution of 720x576 pixels, with each pixel being represented by 8-bits. Thus, when expressed in device units, pixel values range between 0 and 255. The gamma setting of the CCD video camera was adjusted to provide a linear response (gamma=1). GPS time inserters are used to stamp date and time information on every video frame with an accuracy of 0.01 seconds. f/3.3 focal reducers manufactured by Meade are also used in order to increase the area monitored by these devices. The telescopes are tracked at lunar rate, but they are manually recentered when necessary, since perfect tracking of the Moon at the required precision is not feasible with this equipment.

The observations conducted with the 0.36 m and the 0.28 m telescopes were performed without any filter. These provided images in the wavelength range between, approximately, 400 and 1000 nm. However, a NIR filter (model Baader IR-pass) was employed for the camera attached to the 0.24 m telescope. As a result, the images taken by this telescope corresponded to wavelengths raging from 685 to 1000 nm.

Our dates of monitoring did not coincide with the activity period of any major meteor shower. So, the telescopes were aimed to an arbitrary but common area of the lunar disk. Of course, the terminator was avoided in order to avoid an excess of light from the illuminated side of the Moon in the telescopes. For the identification and analysis of impact flashes we have employed the MIDAS software (Madiedo et al. 2010, 2015a).

## 3. OBSERVATIONS

The monitoring campaign on 2015 March 25 was conducted from 19h 15m to 23h 45m UT, with a 5.9 day-old waxing crescent Moon and an illuminated fraction of the lunar disk of about 34 % (Figure 1). The effective monitoring time was 4.5 hours. As a result of this campaign one impact flash was identified (Figure 2). The event took place at 21h 00m 16.80 ± 0.01 s UT and was simultaneously recorded by the 3 telescopes. It lasted 0.18 s according to the observations performed in visible band with the 0.28 m telescope. However, the images taken in the near IR with the 0.24 m telescope revealed that the duration of the event was of around 0.20 s. To calculate the duration of the flash we have employed the MIDAS software. This software tool measured the average luminosity of the area on the lunar surface enclosed by the event, and took into account only those video frames where this luminosity was above the sum of the average luminosity of the same area on the Moon immediately before the event took place and the corresponding standard deviation of that luminosity. It is important to notice that the diameter of the telescope employed to obtain images in the NIR was smaller. This means that the ability of this device to collect light (and hence its sensitivity to record dimmer events) is lower. So the larger duration of the flash in the near-infrared can be explained on the basis of the enhanced contrast that the NIR filter provides between the impact flash and the Earthshine (Figures 2a and 2b).

The main parameters of the impact flash are listed in Table 1. The impactor hit the lunar surface at the selenographic coordinates 11.3 ± 0.1 ºN, 21.6 ± 0.1 ºW, which corresponds to a position close to the northwest wall of crater Copernicus.

## 4. RESULTS AND DISCUSSION
### 4.1. Flash brightness

To estimate the magnitude in our NIR band we have followed the procedure described in Madiedo et al. (2014a). Thus, the brightness of the flash was compared with the brightness of reference stars recorded during the same observing session whose visual magnitude is known. This was done by using the stars magnitudes in I-band, which is the closest band to our filter in the Bessel Cousins photometric system. Eight calibration stars with similar airmass and apparent brightness to that of the flash were employed. Even though the transmission of our filter does not exactly match the shape and width of the standard I filter, it is reasonably close, given that the sensitivity of the CCD camera quickly decays in this wavelength range. The response of the Johnson-Cousins I filter transmission and the Baader IR-pass filter transmission convolved with the spectral response of the camera CCD is shown in (Figure 3). From this plot we have obtained that the effective wavelength for the Johnson-Cousins I filter and the Baader IR-pass filter is similar (798.0 and 761.5 nm, respectively). As a result of this procedure we have obtained that the magnitude of the flash in the near infrared was $m_I = 5.1 \pm 0.3$. The V-band magnitude $m_V$ of the event was also obtained by employing reference stars recorded during the same observing session with known magnitude in V band. In previous works we have never used colour terms in the flux to V-magnitude conversion equations, mainly because we did not know the colour of the impact flashes. But since in this case observations in the NIR are available, by using the information from the rough V-I colour determined from our data we can do that, at least approximately. Given that the sensitivity of the CCD camera is not solely centered in the V band, and we did not use a V-band filter, NIR radiation can enter the detector and so it is reasonable to expect that this effect should be taken into account. In fact, we believe that a large fraction of the flux in our unfiltered observations includes NIR flux. Thus color term corrections seem necessary. The V-band magnitude was calculated by using the standard relationship

$$m_V = m_{V_o} - 2.5\log(S) + K_I(m_V - m_I) - KX \qquad (1)$$

where $m_{V_o}$ is the zero-point for the V filter, S the measured signal, $K_I$ the colour term transformation coefficient, K the extinction coefficient, and X the airmass. We employed calibration stars, with known $m_V$ and $m_I$, to obtain the values of $K_I$, $m_{V_o}$ and KX. These stars were observed with our telescopes to measure their corresponding signal S. We considered eight calibration stars with the same airmass as the impact flash, so that in this calibration we simultaneously determined $m_{V_o}$ and KX in a single constant by performing a least-squares fit. The value of $K_I$ resulting from this calibration is 0.18, and the peak magnitude of the flash in V band yields 7.3 ± 0.2. We have assumed a typical error for $K_I$ of about 30 %. This error was provided by the above-mentioned fit. It is important to point out that we should have used a similar transformation equation for the I magnitude as we did for the V magnitude in Equation (1). But unfortunately we could not determine color correction terms for the I filter. For this reason, the uncertainty for the flash magnitude in V band is lower than the uncertainty in I band.

### 4.2. Meteoroid source

It is a well-known fact that it is not possible to unambiguously establish the source of the impactors from the monitoring of lunar impact flashes. However, it is possible to determine the most likely source of these projectiles by calculating the probability that an impact flash is associated to a given meteoroid source (Madiedo et al. 2015a,b). Following this approach, the impact flash can be linked to the meteoroid source that provides the highest value for this probability.

At the time of the impact flash detection no major meteor shower was active on Earth. Thus, we have considered that the event could be produced either by a sporadic meteoroid or by a particle belonging to a minor meteoroid stream. From March 24-26 our meteor observing stations (Madiedo and Trigo-Rodríguez 2008; Madiedo et al. 2013; Madiedo 2014) recorded activity from the Virginids (VIR) and the γ-Normids (GNO), both with a zenithal hourly ZHR of below 1 meteor $h^{-1}$. The impact geometry of both streams was compatible with that of the impact flash discussed here. So, we have assumed that the impactor could be linked to one of these two meteoroid streams or to the sporadic background. The corresponding association probabilities, labelled by $p_{VIR}$, $p_{GNO}$ and $p_{SPO}$, respectively, were calculated according to the methods described in Madiedo et al. (2015a,b).

Thus, by using Eq. (15) in Madiedo et al. (2015b) with ZHR = 1 meteors $h^{-1}$ for both streams and by assuming an hourly rate of 10 meteors $h^{-1}$ for sporadics (Dubietis and Arlt 2010), $p_{VIR}$ and $p_{GNO}$ yield 0.04 and 0.08, respectively. For the calculation of these probabilities we have considered a population index r = 3.0 for the Virginids, r = 2.4 for the γ-Normids, and r = 3.0 for sporadics. An average value of 17 km $s^{-1}$ (Ortiz et al. 1999) has been assumed for the impact velocity on the Moon of sporadic meteoroids. For the Virginids and the γ-Normids we have taken these impact velocities as 30 and 56 km $s^{-1}$, respectively. An impact angle of 45º with respect to the local vertical was considered for sporadic events, while according to the impact geometry this angle would be of about 40.2º and 46.2º for the Virginids and the γ-Normids, respectively. These results show that the most likely scenario is that the meteoroid was associated with the sporadic background, since the largest probability parameter is obtained for this source, with $p_{SPO}$ = 1 - $p_{VIR}$ - $p_{GNO}$ = 0.88.

### 4.3. Emission efficiency in the near infrared and impactor mass

The lightcurve of the impact flash in visible and NIR bands is shown in Figure 4. As can be noticed, and within the time resolution provided by our cameras, the flash peaks at the same instant in both bands, but it remains brighter in the near-infrared during the whole duration of the event.

The radiated energy on the Moon corresponding to an impact flash with a magnitude m can be obtained by integrating the radiated power P defined by the following equation:

$$P = 3.75 \cdot 10^{-8} \cdot 10^{(-m/2.5)} f\pi\Delta\lambda R^2 \qquad (2)$$

where P is given in watts, $3.75 \cdot 10^{-8}$ is the flux density in V band of a magnitude 0 star in $Wm^{-2}\mu m^{-1}$, $\Delta\lambda$ is the wavelength range, expressed in μm, in which P is calculated, R is the Earth-Moon distance at the instant of the meteoroid impact in m, and f is a dimensionless factor that measures the degree of anisotropy of light emission. Thus, for those impacts where light is isotropically emitted from the surface of the Moon f = 2, while f = 4 if light is emitted from a very high altitude above the lunar surface. In this analysis we have assumed f = 2.

Thus, by integrating Eq. (2) we have obtained the radiated energy on the Moon in the luminous range (the CCD sensitivity range) by employing the visual magnitude $m_V$ = 7.3 and a passband $\Delta\lambda$ = 0.5 μm (Bellot Rubio et al. 2000a,b; Ortiz et al. 2000). Also the corresponding energy radiated in the near-infrared by employing m = 5.1 and $\Delta\lambda$ = 0.315 μm. Note that the zero magnitude flux density for I band is $9.76 \cdot 10^{-9}$ $Wm^{-2}\mu m^{-1}$ according to Bessel (1998) and not $3.75 \cdot 10^{-8}$ $Wm^{-2}\mu m^{-1}$. The Earth-Moon distance at the instant of the impact was R= 389134.5 km. According to our calculations these energies yield $E_V$ = $1.46 \cdot 10^6$ J and $E_I$ = $1.44 \cdot 10^6$ J, respectively. This smaller value mainly comes from the smaller bandpass of our near-infrared observations. If we used a normalized passband of 0.5 microns for the two cases, or use the flux density, the energy emitted in a normalized passband in the infrared would be 0.5/0.315 times larger or $E_I = 2.28 \cdot 10^6$ J. Hence there is a significant increase of the efficiency in the near-infrared compared to that in the visible.

To estimate the kinetic energy of the impactor, the meteoroid mass and the emission efficiency in the infrared, we have assumed that the luminous efficiency in V band for sporadic impact flashes is $2 \cdot 10^{-3}$ (Ortiz et al. 2006, 2015). However, it must be taken into account that this efficiency in V band was obtained by employing f = 3 in Eq. (2). Since in this work we have taken f = 2, we must multiply this efficiency by a factor 3/2, so that the assumed luminous efficiency in V band for the event considered here yields $\eta_V = 3 \cdot 10^{-3}$. The kinetic energy $E_k$ of the impactor is then given by the equation

$$E_k = E_V/\eta_V \quad\quad\quad\quad\quad\quad (3)$$

and yields $E_k = 4.86 \cdot 10^8$ J.

Once the kinetic energy is known, the emission efficiency in our NIR band $\eta_{NIR}$ can be estimated from

$$\eta_I = E_I/E_k \tag{4}$$

This parameter yields $\eta_I = 4.7 \cdot 10^{-3}$. Nevertheless, this result depends critically on the value adopted for $\eta_V$. For $\eta_V = 5 \cdot 10^{-4}$ and $\eta_V = 5 \cdot 10^{-3}$ the kinetic energy of the impactor yields $2.91 \cdot 10^9$ J and $2.91 \cdot 10^8$ J, respectively, while the resulting emission efficiency in the infrared yields $7.8 \cdot 10^{-4}$ and $7.8 \cdot 10^{-3}$, respectively. According to this calculation, the emission efficiency for sporadic events in this spectral band is higher than in V band by a factor of about 56%. Notice that this factor does not depend on the value adopted for $\eta_V$. This factor shows that presumably a large part of the electromagnetic energy released as a consequence of the impact is radiated in the infrared.

For an impact velocity V for sporadics of 17 km s$^{-1}$ the kinetic energy $E_k$ calculated by assuming $\eta_V = 3 \cdot 10^{-3}$ corresponds to an impactor mass M = 3.4 ± 0.3 kg. To obtain the impactor size we have considered a bulk density of 1.8 g cm$^{-3}$ for sporadic meteoroids (Babadzhanov and Kokhirova 2009). According to this assumption, the impactor diameter would be $D_P$ = 15.3 ± 0.4 cm. For $\eta_V = 5 \cdot 10^{-4}$ and $\eta_V = 5 \cdot 10^{-3}$ this calculation yields M = 20 ± 3 kg and M = 2.0 ± 0.3 kg, respectively, and the corresponding meteoroid diameter yields $D_P$ = 27 ± 1 cm and $D_P$ = 12.8 ± 0.6 cm, respectively.

### 4.4. Impact plume temperature

From the magnitudes shown in Figure 4, the evolution with time of the energy flux density measured on Earth for both V and I bands (denoted by $F_V$ and $F_I$, respectively) can be obtained by employing the following relationships:

$$F_V = 3.75 \cdot 10^{-8} \cdot 10^{(-m_V/2.5)} \tag{5}$$

$$F_I = 9.76 \cdot 10^{-9} \cdot 10^{(-m_I/2.5)} \tag{6}$$

where, as mentioned above, $3.75 \cdot 10^{-8}$ and $9.76 \cdot 10^{-9}$ are the irradiance of a magnitude 0 star in $Wm^{-2}\mu m^{-1}$ for V and I bands, respectively (Bessel 1998). If we assume that the intensity distribution of the impact flash follows Planck's law, the ratio of these flux densities must satisfy the equation

$$\frac{F_V}{F_I} = \left(\frac{\lambda_I}{\lambda_V}\right)^5 \frac{e^{hc/k\lambda_V T} - 1}{e^{hc/k\lambda_I T} - 1} \tag{7}$$

In this relationship $\lambda_V = 0.55$ μm and $\lambda_I = 0.798$ μm are the effective wavelengths for V and I bands, respectively (Bessel 2005), h is Planck's constant, c the speed of light in vacuum, and k is Boltzmann's constant. The evolution with time of the impact plume temperature T estimated by solving Eq. (7) is shown in Figure 5. This plot shows that this temperature reached a maximum value of around 4000 K at the beginning of the flash, and then after a sudden decrease to around 3200 K it remains practically constant for around 0.1 s. This suggests that during this phase the condensation process gives rise to equilibrium in the impact plume, so that the temperature remains constant as a consequence of the release of evaporation energy (Nemtchinov 1998). After that time the plume temperature slowly decreases to a final value of ~2900 K by the end of the event.

### 4.5. Crater size

The results obtained from the analysis of the crater resulting from the impact of the meteoroid on the lunar surface are summarized in Table 2. We have employed the crater-scaling equation for the Moon given by Gault, which is valid for craters with a diameter of up to about 100 meters in loose soil or regolith (Gault 1974, Melosh 1989):

$$D = 0.25 \rho_p^{1/6} \rho_t^{-0.5} E_k^{0.29} (\sin \theta)^{1/3} \tag{8}$$

Magnitudes in this relationship must be entered in mks units. D is the crater diameter, $E_k$ the kinetic energy of the impactor, $\rho_p$ and $\rho_t$ are the impactor and target bulk densities, respectively, and $\theta$ is the impact angle with respect to the horizontal. Since this angle is unknown for sporadic meteoroids (the most likely source of the impact flash), we have considered $\theta = 45º$, which is the value of the most likely impact angle. The rim-to-rim crater diameter derived from Eq. (8), with the value of $E_k$ obtained with $\eta_V = 3 \cdot 10^{-3}$, yields $D = 6.4 \pm 0.2$ m for an impactor bulk density of 1.8 g cm$^{-3}$. For the target bulk density we have taken $\rho_t$ = 1.6 g cm$^{-3}$. As an alternative formula to obtain the rim-to-rim diameter of this crater we have employed the following equation (Holsapple 1993):

$$D = 2.6 K_r \left[ \frac{\pi_v m}{\rho_t} \right]^{1/3}. \qquad (9)$$

where the adimensional factor $\pi_v$ is obtained from

$$\pi_v = K_1 \left[ \left( \frac{ga}{(V \sin(\theta))^2} \right) \left( \frac{\rho_t}{\rho_P} \right)^{\frac{6v-2-\mu}{3\mu}} + \left[ K_2 \left( \frac{Y}{\rho_t (V \sin(\theta))^2} \right) \left( \frac{\rho_t}{\rho_P} \right)^{\frac{6v-2}{3\mu}} \right]^{\frac{2+\mu}{2}} \right]^{\frac{-3\mu}{2+\mu}} \qquad (10)$$

In Equations (9) and (10) physical quantities are entered in mks units, with $K_1$=0.2, $K_2$=0.75, $K_r$=1.1, $\mu$=0.4, $\nu$=0.333 and Y = 1000 Pa. V, a, and m are the impactor velocity, radius and mass, respectively. The gravity on the lunar surface is g = 0.162 m s$^{-2}$. By using Equations (9) and (10) one obtains D = 6.8 ± 0.2 m for $\rho_p$ = 1.8 g cm$^{-3}$ and V= 17 km s$^{-1}$.

It must be also taken into account that despite $\rho_p$ = 1.8 g cm$^{-3}$ has been adopted above, the density of sporadic meteoroids can range from 0.3 g cm$^{-3}$ (the density of soft cometary materials) to 3.7 g cm$^{-3}$ (the density of ordinary chondrites). For these densities the diameter of this crater calculated from Eq. (8) would be 4.8 ± 0.1 m and 7.3 ± 0.2 m, respectively.

The other two possible sources of the projectile are the Virginids, which follow an asteroidal orbit, and the γ-Normids, which have a cometary nature (Jenniskens 2006). So, for the Virginids we have considered projectile densities ranging between 2.4 g cm$^{-3}$ (the corresponding to carbonaceous chondrites) and 3.7 g cm$^{-3}$ (the density of ordinary chondrites). And for the γ-Normids we have assumed that this density can range between 0.3 g cm$^{-3}$ (the density of soft cometary materials) and 1.8 g cm$^{-3}$. As mentioned above, the impact velocity for the Virginids is 30 km s$^{-1}$, with an impact angle of 40.2º with respect to the local vertical. And for the γ-Normids the impact velocity is 56 km s$^{-1}$, with an impact angle with respect to the local vertical of 46.2º. According to this, the Gault model yields for the Virginids a crater diameter ranging from 6.9 ± 0.2 m (for $\rho_p$ = 2.4 g cm$^{-3}$) to 7.4 ± 0.2 m (for $\rho_p$ = 3.7 g cm$^{-3}$), and for the γ-Normids this diameter would range from 4.7 ± 0.1 m (for $\rho_p$ = 0.3 g cm$^{-3}$) to 6.4 ± 0.2 m (for $\rho_p$ = 1.8 g cm$^{-3}$).

Despite this crater is too small to be directly observed with instruments from Earth, it could be observed by means of probes orbiting the Moon (with before and after images taken under similar illumination conditions), such as for instance the Lunar Reconnaissance Orbiter (LRO) (Robinson 2015). In this way, the actual crater size could be compared with the predicted diameter in order to test the validity of the assumptions employed in this analysis.

## 5. CONCLUSIONS

We have analyzed a lunar impact flash recorded on 25 March 2015 in both visual and NIR bands. Despite the fact that estimations of the temperature of lunar impact flashes have been published by other team during the refereeing process of our manuscript (Bonanos et al. 2018), our result still forms the first measurement of the temperature of a telescopic lunar impact flash (Madiedo and Ortiz 2016, 2018). The peak visual magnitude of this event was 7.3 ± 0.2 and the peak magnitude in I band was 5.1 ± 0.3. Our calculations show that the most likely origin of the source meteoroid is the sporadic background, with a probability of 88 %. By assuming a luminous efficiency in V-band of $3 \cdot 10^{-3}$, the impactor mass for this sporadic projectile yields 3.4 ± 0.3 kg. And the estimated diameter D of the resulting crater would range, according to the Gault model, from 4.8 ± 0.1 m (for $\rho_p$ = 0.3 g cm$^{-3}$) to 7.3 ± 0.2 m (for $\rho_p$ = 3.7 g cm$^{-3}$), with D = 6.8 ± 0.2 m for $\rho_p$ = 1.8 g cm$^{-3}$. This fresh crater could be observed by means of probes orbiting the Moon, such as LRO.

The emission efficiency in the NIR for this sporadic event has been also inferred. The value of this parameter yields $4.7 \cdot 10^{-3}$, which is higher, by around 56%, than the luminous efficiency in visible band. This shows that presumably a large part of the electromagnetic energy radiated as a consequence of the impact is emitted in the near infrared. The temperature of the impact plume, which has been obtained from the energy flux densities for V and I bands, is of around 3200 K during most of the light curve, which suggests that condensation gives rise to equilibrium in the impact plume during this stage. But the very initial flash is even hotter (about 4000 K), and this temperature decreases to ~2900 K by the end of the event.

## ACKNOWLEDGEMENTS

The authors acknowledge support from projects AYA2014-61357-EXP (MINECO), AYA2015-68646-P (MINECO/FEDER), Proyecto de Excelencia de la Junta de Andalucía, J.A. 2012-FQM1776, and from FEDER.

**TABLES**

| | |
|---|---|
| Date and time | 2015 March 25 at 21h 00m 16.80 ± 0.01s UT |
| Peak brightness (magnitude) | 7.3±0.2 in V band; 5.1±0.3 in I-band |
| Selenographic coordinates | Lat.: 11.3±0.1 °N, Lon.: 21.6±0.1 °W |
| Duration (s) | 0.18 (V band); 0.20 (NIR) |

Table 1. Characteristics of the impact flash discussed in this work

| Meteoroid source | Model | Impact angle (°) | Meteoroid Density (g cm$^{-3}$) | Meteoroid Mass (kg) | Impact Velocity (km s$^{-1}$) | Crater Diameter (m) |
|---|---|---|---|---|---|---|
| Sporadic | Gault | 45 | 0.3 | 3.4±0.3 | 17 | 4.8±0.1 |
| | Gault | 45 | 1.8 | 3.4±0.3 | 17 | 6.4±0.2 |
| | Gault | 45 | 3.7 | 3.4±0.3 | 17 | 7.3±0.2 |
| | Holsapple | 45 | 0.3 | 3.4±0.3 | 17 | 6.8±0.2 |
| | Holsapple | 45 | 1.8 | 3.4±0.3 | 17 | 6.8±0.2 |
| | Holsapple | 45 | 3.7 | 3.4±0.3 | 17 | 6.8±0.2 |
| Virginids | Gault | 40.2 | 2.4 | 1.1±0.1 | 30 | 6.9±0.2 |
| | Gault | 40.2 | 3.7 | 1.1±0.1 | 30 | 7.4±0.2 |
| | Holsapple | 40.2 | 2.4 | 1.1±0.1 | 30 | 6.2±0.2 |
| | Holsapple | 40.2 | 3.7 | 1.1±0.1 | 30 | 6.2±0.2 |
| γ-Normids | Gault | 46.2 | 0.3 | 0.31±0.05 | 56 | 4.7±0.1 |
| | Gault | 46.2 | 1.8 | 0.31±0.05 | 56 | 6.4±0.2 |
| | Holsapple | 46.2 | 0.3 | 0.31±0.05 | 56 | 5.2±0.2 |
| | Holsapple | 46.2 | 1.8 | 0.31±0.05 | 56 | 5.2±0.2 |

Table 2. Rim-to-rim crater diameter predicted by the Gault and the Holsapple models, by assuming $\eta_V = 3 \cdot 10^{-3}$. The impact angle is measured with respect to the local vertical.

**FIGURES**

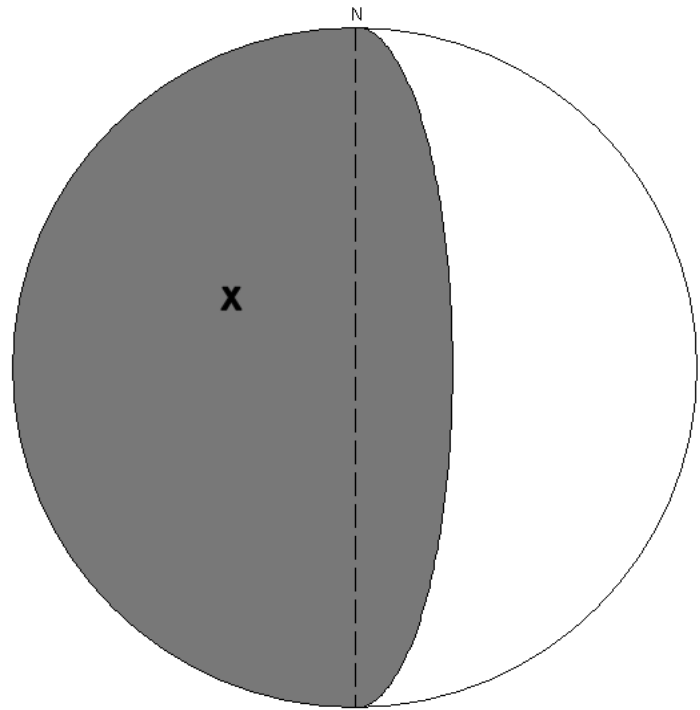

Figure 1. The lunar disk as seen from Earth on 2015 March 25. White region: area illuminated by the Sun. Gray region: night side. Cross: impact flash position.

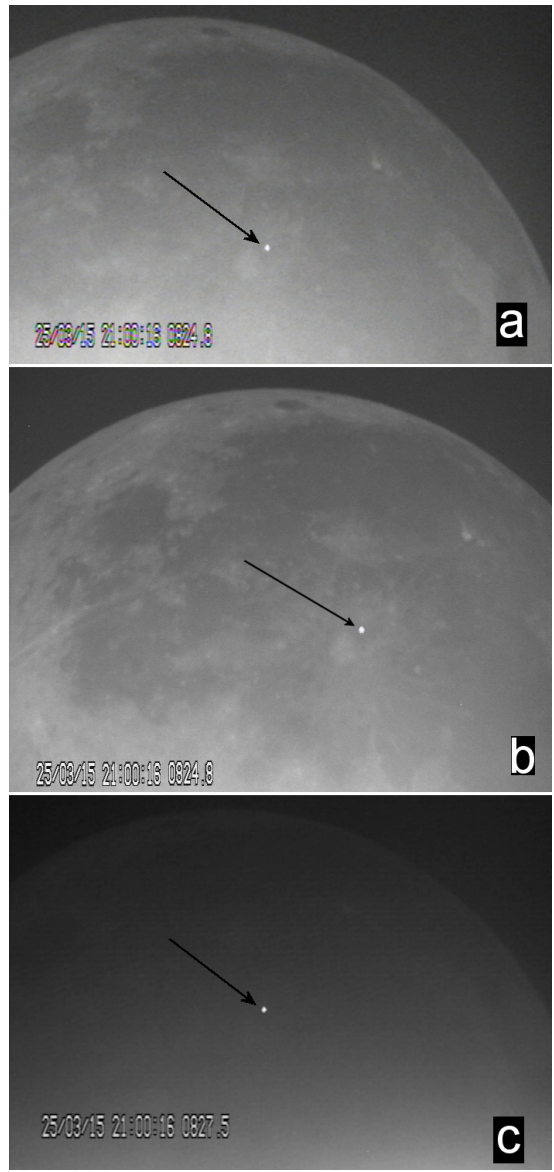

Figure 2. Impact flash detected from Sevilla on 2015 March 25 as recorded by the 0.28 cm (a), the 0.36 m (b) and the 0.24 cm (c) SC telescopes

operating at this observatory. Images (a) and (b) were recorded with unfiltered cameras. Image (c) was obtained by means of a NIR filter.

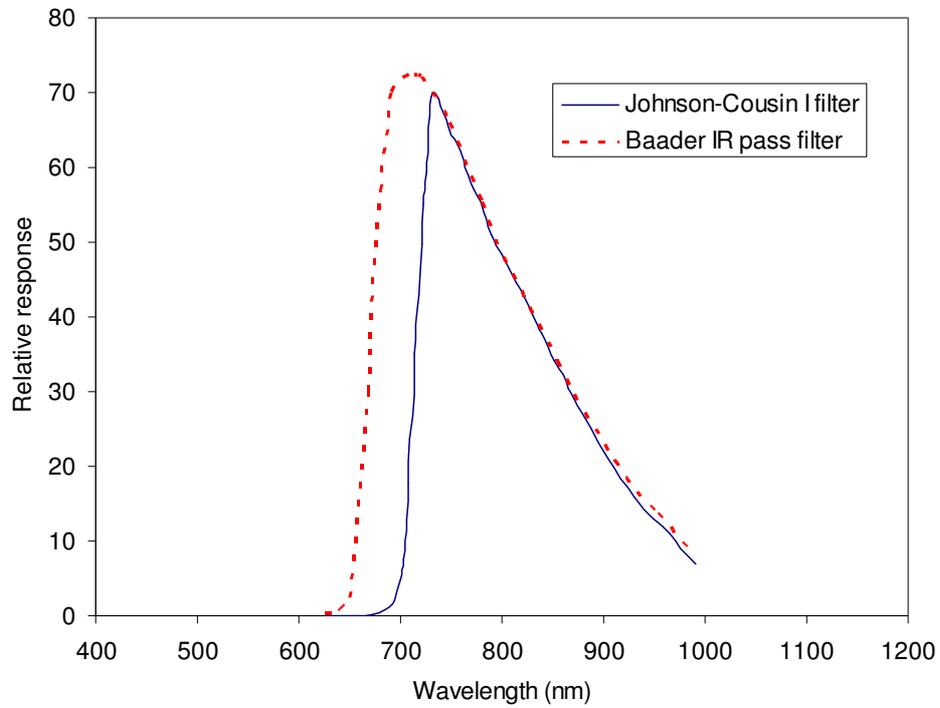

Figure 3. Response of the Johnson-Cousins I filter transmission and the Baader IR-pass filter transmission convolved with the spectral response of the camera CCD.

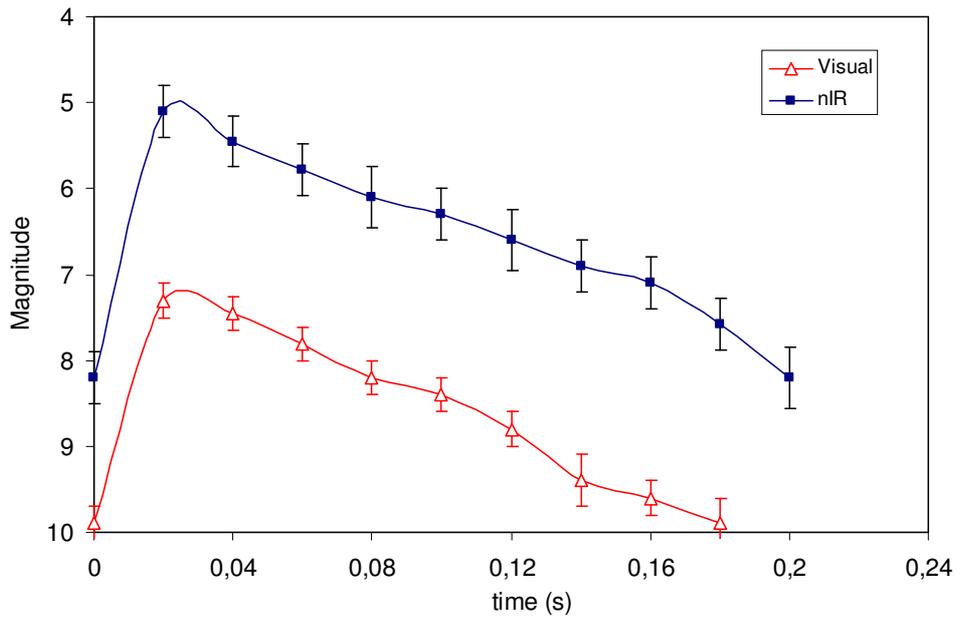

Figure 4. Lightcurve of the impact flash in visible light and in the NIR. Visual data correspond to the observation performed with the 0.36 m telescope.

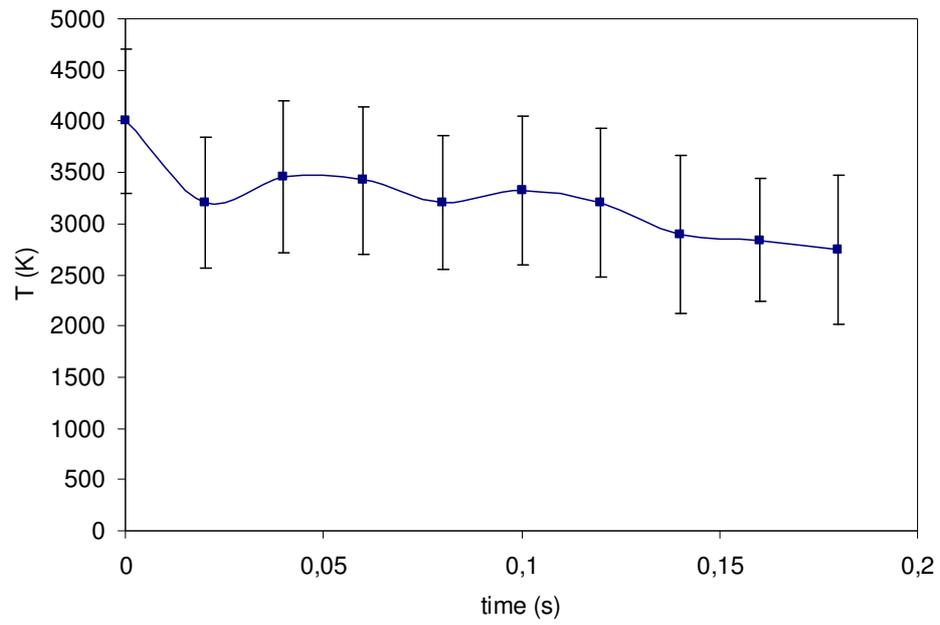

Figure 5. Evolution with time of the temperature of the impact plume, with a spline curve fitted through the points.